# Predicting Human Lifespan Limits

Byung Mook Weon[1,2] and Jung Ho Je[2]

[1] *Department of Physics, School of Engineering and Applied Sciences, Harvard University, Cambridge, Massachusetts 02138, USA.*

[2] *X-ray Imaging Center, Department of Materials Science and Engineering, Pohang University of Science and Technology, Pohang 790-784, Korea.*

**ABSTRACT**

Recent discoveries show steady improvements in life expectancy during modern decades. Does this support that humans continue to live longer in future? We recently put forward the maximum survival tendency, as found in survival curves of industrialized countries, which is described by extended Weibull model with age-dependent stretched exponent. The maximum survival tendency suggests that human survival dynamics may possess its intrinsic limit, beyond which survival is inevitably forbidden. Based on such tendency, we develop the model and explore the patterns in the maximum lifespan limits from industrialized countries during recent three decades. This analysis strategy is simple and useful to interpret the complicated human survival dynamics.

**Keywords:** Human lifespan; Maximum survival tendency; Lifespan limit

## 1. INTRODUCTION

Humans live longer now. Indeed the life expectancy and the observed maximum age at death have significantly increased during recent decades [1−5]. Such increase is mainly attributable to non-biological aspects such as the intricate interplay of advances in income, nutrition, education, sanitation, and medicine [6,7]. Biologists and gerontologists are hunting for a variety of useful ways to prolong life in animals, including mice and worms [8]. Their research suggests that human lifespan may be remarkably pliable [8]. Can the strategies for animals help humans live longer? So far it is not practical, useful, or ethical to extend healthy life merely by modifying human genes [9] or by restricting food intake [10]. The theoretical maximum lifespan (called ω) in humans is still a subject of considerable debate [6] and the life extension is one of the great challenges in the 21st century [8]. Many scientists believe that human lifespan has an inherent upper limit, although they disagree on whether it is 85 or 100 or 150 [8]. The maximum human lifespan is generally postulated to be around 125 years [7,11,12], whereas the record of the oldest ages at death is increasing today [4]. Conventional analysis or theoretical model has not yet come up with a plausible explanation for this disagreement.

Recently, based on extended Weibull model with age-dependent stretched exponent [13,14], we suggested a mathematical model for human survival dynamics, S(x) =

exp(–(x/α)^β(x)), which denotes survival probability with characteristic life α and age-dependent exponent β(x), and showed maximum survival tendency, dS(x)/dx → 0 [15]. In this study, we further develop the model and explore the dynamic patterns with year and country in predicting human lifespan limits (ω) for industrialized countries during recent three decades: ω = 0.458q + 54.241 where the upper x-intercept $q = h + (k/p)^{1/2}$ for the quadratic model $β(x) = –p(x – h)^2 + k$ (where p, h, and k are variable with year and country). We aim to examine the lifespan puzzle—whether human lifespan is approaching a limit or not. Our analysis strategy has practical implications for aging research in biology, medicine, statistics, economy, public policy, and culture.

## 2. METHODS

We examine the survival dynamics of Sweden female's survival curves during recent three decades, from 1977 to 2007. The reliable demographic data were taken from the periodic life tables (1x1) at the Human Mortality Database (http://www.mortality.org). We analyze the survival curves by using a general expression of human survival probability (S(x)) as a function of age (x) [15]:

$$S(x) = \exp(-(x/\alpha)^{\beta(x)})$$

Here, the characteristic life (α) corresponds to the specific age of S(α) = exp(–1) and the age-dependent stretched exponent (or beta function, β(x)) reflects the flexibility of the survival curve [15]. The survival function allows the cumulative hazard function M(x) (= –log S(x)) on a restricted range. The breakdown of the positivity of the hazard function m(x) (= ∂M(x)/∂x) enables us to estimate a maximum limitation of human lifespan. Intuitively, our survival model approximates the Gompertz model [16] with a linear expression for β(x) as well as the Weibull model [17] with a constant β(x) through an approximation of 'log m(x) ∝ β(x)'.

**Figure 1** illustrates the evolution of the beta function β(x), which is a pure mathematical conversion of the survival probability S(x), for Swedish females from 1977 to 2007. The smooth survival data points above 94-97 years were chosen for modeling the beta function (solid lines). Here the discontinuities of the beta curves (dashed lines) near the characteristic lives (around 85-90 years) are due to the mathematical feature of the suggested model [15]. Apparently the curvatures of the beta curves seem to become more "negative" at the highest ages and the vertex points move upward year by year. Such trends directly connote the emergence of the "maximum survival tendency" [15].

In principle, the age-dependent beta function originates from the "maximum survival tendency", which is a fundamental biological feature of human survival dynamics by minimizing its death rate (dS(x)/dx → 0) [15]. The maximum survival tendency is characterized as a "negative" slope of the beta function as $d^2β(x)/dx^2 < 0$ for the phase of x > α.

We find that a quadratic model, $β(x) = –β_0 + β_1x – β_2x^2$ (where $β_0, β_1, β_2 > 0$), is appropriate to describe the maximum survival tendency from the modern survival curves for the highest ages (for x > α), as marked by the solid lines in **Fig. 1**. This agrees to our previous observation [15]. In this study, we modify the quadratic model for β(x) as:

$$\beta(\mathrm{x}) = -p(x-h)^2 + k$$

Here the coefficient p (= $\beta_2$ = (–1/2) $d^2\beta(x)/dx^2$) denotes the curvature of the quadratic curve and the vertex point v(h, k) indicates the maximum value of the quadratic curve. The curvature and the vertex point give an upper x-intercept (**Fig. 1**), as can be defined as the "q" point:

$$q = h + (k/p)^{1/2}$$

The quadratic beta function based on the maximum survival tendency can be entirely described by quantifying the v(h, k) and the q points.

### 3. RESULTS

The intrinsic definition of β(x) and S(x) leads to a mathematical limitation of the survival age, beyond which none can be alive. The theoretical limitation of the maximum lifespan (ω) is determined at the specific age of β(x) = f(x) as seen in **Fig. 1**. Here f(x) is the mathematical constraint of β(x) as defined as f(x) = –xln(x/α) dβ(x)/dx [15]. This feature suggests that the ω value can be found between the v(h, k) and the q points.

We observe the evolution of the quadratic beta functions from the survival curves of Sweden females, as seen in **Fig. 2**. The p and the k parameters linearly increase by period (P): p = 6.8897×10$^{-5}$ P – 0.1346 and k = 6.389×10$^{-2}$ P – 116.581. By contrary, the h parameter does not significantly change (average ~ 95.482 years), obviously since 1985, as marked by the gray area in **Fig. 2**. The linear increases of the p and the k parameters indicate that the curvature of the quadratic function becomes more negative and the vertex point moves upward from 1977 to 2007. Our model suggests that the upper x-intercept (q-value) may significantly decrease by period, following the scaling of (k/p)$^{1/2}$. The obtained q values (squares) well follow the trend line (solid line) which is estimated from the p and the k parameters. The parameter estimation for Sweden females is summarized at **Table 1**. The correlation coefficients ($r^2$) between data and model are higher than 0.994, suggesting the feasibility of the model. As a result, we see in **Fig. 2** that the q parameters gradually decrease by period in Sweden female's life tables during recent three decades.

The most interesting observation is that the maximum lifespan limits (ω) have a linear relationship with the upper x-intercept (q) parameter, as clearly seen in **Fig. 3**. The high linearity between the ω and the q values is found for both cases of Sweden females (between 1977 and 2007) and modern industrialized countries (Austria, Belgium, Bulgaria, Canada, Czech, Denmark, England, Estonia, Finland, France, West Germany, Hungary, Iceland, Ireland, Italy, Japan, Latvia, Lithuania, Netherlands, Norway, Poland, Russia, Scotland, Slovenia, Spain, Sweden, Switzerland, and USA; for females between 2005 and 2007). The parameter estimation for modern industrialized countries is summarized at **Table 2**. The correlation coefficients ($r^2$) between data and model are higher than 0.954, suggesting again the feasibility of the model. Interestingly, we find in **Fig. 3** that the three-decade variation of Sweden female (squares) is similar to the national variation of the other countries (circles). This similarity suggests that the dataset of Sweden female can be indeed "representative" for human survival tendency as suggested [4]. In **Fig. 3**, we see that the ω values for all the datasets linearly decrease as the q values decrease ($r^2 = 0.9445$):

$$\omega = 0.458q + 54.241$$

It is interesting that the ω values shift toward ~125 years (close squares) for Sweden females during the latest decade from 1997 to 2007.

## 4. DISCUSSION

The overall evolutions of the q values (**Fig. 2**) and ω values (**Fig. 3**) suggest that the human lifespan would be reaching an upper limit. Our study implies that the observed maximum lifespan limit is able to continue to climb until it encounters a theoretical forbidden barrier of human lifespan, as suggested [18]. The life-extension strategies such as aggressive anti-aging therapies may allow more people to reach the limit of the natural human lifespan and thus the period of disease or senescence will be compressed against the natural barrier at the end of life, as expected [19,20].

The lifespan limit estimation may support current aging theories that presume the existence of the biological limit to human lifespan [21−23]. Based on our estimation, it is predictable that many countries will face increasing issues of aging populations, age-related diseases, and healthcare costs [8]. The rise of human longevity will accelerate the population growth rate [24] and probably the steady rise in the achieved maximum lifespan [4] or the life expectancy [25] will reduce in the coming half century. The forthcoming trends may cause an ethical issue on fair distribution of healthcare resources [26]. Aging research requires new approaches to figure out the complex biology of aging [27].

The feasibility of the model is further obtained through a mathematical verification [28]: our data exist between $0.4 < (\omega/\alpha)\ln(\omega/\alpha) < 0.8$, which are consistent with their mathematical expectation between 0.410986 and 0.829297. Another verification is obtained from mortality patterns, which are defined as $\mu(x) = -d\ln S(x)/dx = d[(x/\alpha)^{\beta(x)}]/dx$ or $\int \mu(x) = (x/\alpha)^{\beta(x)}$. For simplification one defines $\delta(x) = \mu(x)/\int \mu(x) = \beta(x)/x + \ln(x/\alpha)d\beta(x)/dx$. The point where the mortality curve starts to decline is obtained from $\delta(x)^2 + d\delta(x)/dx = 0$ by solving $d\mu(x)/dx = 0$. This condition can be tested by graphical analysis or numerical simulation. For instance, taking the parameters: $\alpha = 88.57223$ years, $\beta_0 = 26.32347$, $\beta_1 = 0.78810$, and $\beta_2 = 0.00409$ from 2007 Swedish female's data (**Table 1**), we obtain the point as ~111 years. Above that point, the mortality curve decreases and eventually reaches zero at the maximum lifespan ~122.86 years. With the quadratic pattern of $\beta(x)$, the mortality pattern tends to decrease after a plateau and ultimately approach zero, well matching typical human mortality patterns. These results show the feasibility of the model.

## 5. CONCLUSION

To conclude, we develop a human survival dynamics model as $S(x) = \exp(-(x/\alpha)^{\beta(x)})$ with $\beta(x) = -p(x - h)^2 + k$ (where p, h, and k are variable with year and country), and explore the pattern of the parameters, $q = h + (k/p)^{1/2}$ and $\omega = 0.458q + 54.241$, which are useful in predicting human lifespan limits (ω). We show generality and feasibility of the model for modern industrialized countries during recent three decades. Based on statistical approach, we suggest that human lifespan is approaching a true limit around 125 years. This estimate may shed light on the central puzzle in aging research: whether biological lifespan limits exist or not. Our model and prediction method would be useful to assess the complicated human survival dynamics [29], which would be essential to study on biology, medicine, statistics, economy, public policy, and culture.

## ACKNOWLEDGEMENTS
We are grateful to the Human Mortality Database (http://www.mortality.org) for allowing anyone to access the demographic data for research. This work was supported by the Creative Research Initiatives (Functional X-ray Imaging) of MEST/KOSEF.

**Figures and Tables**

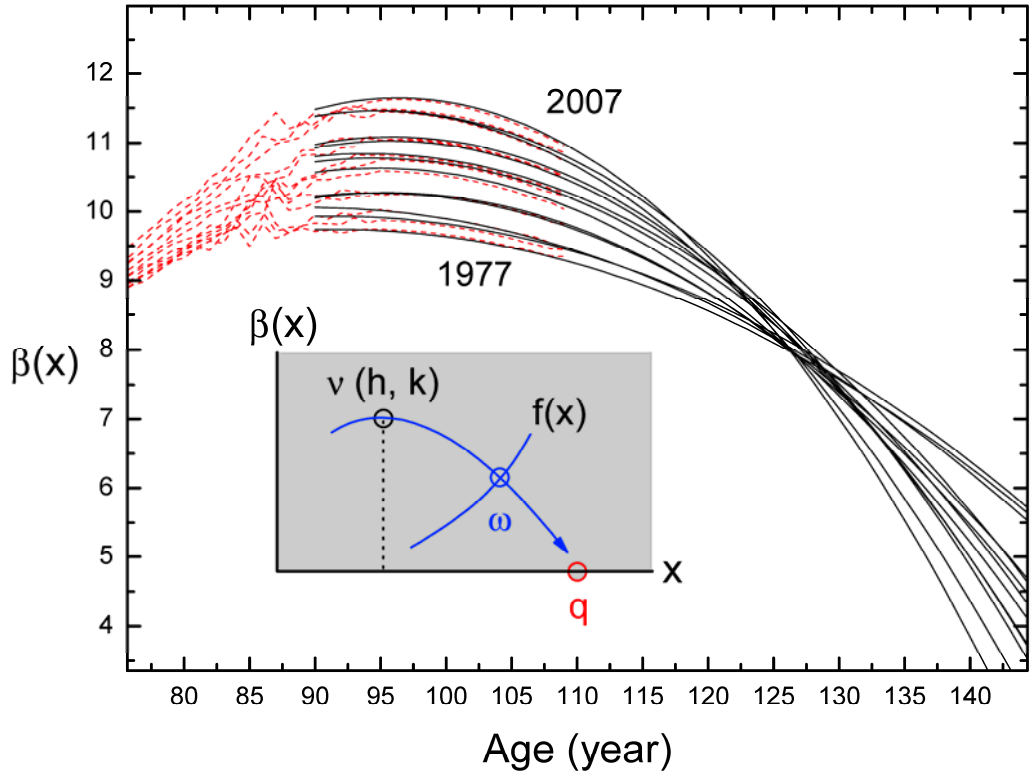

**Figure 1:** The evolution of Sweden female's survival curves (dashed lines) from 1977 to 2007. The beta function β(x) is plotted as a conversion of the survival probability, $S(x) = \exp(-(x/\alpha)^{\beta(x)})$, where the characteristic life α corresponds to the specific age of $S(\alpha) = \exp(-1)$ and the age-dependent beta function β(x) reflects the flexibility of the survival curve. Apparently the β(x) curvature becomes more negative and the vertex point moves upward year by year from 1977 to 2007. The inset describes that the maximum lifespan (ω) is determined at the specific age of β(x) = f(x), which is defined as $f(x) = -x\ln(x/\alpha)\, d\beta(x)/dx$. This feature suggests that the ω value can be found between the vertex point "v(h, k)" and the upper x-intercept "q" point.

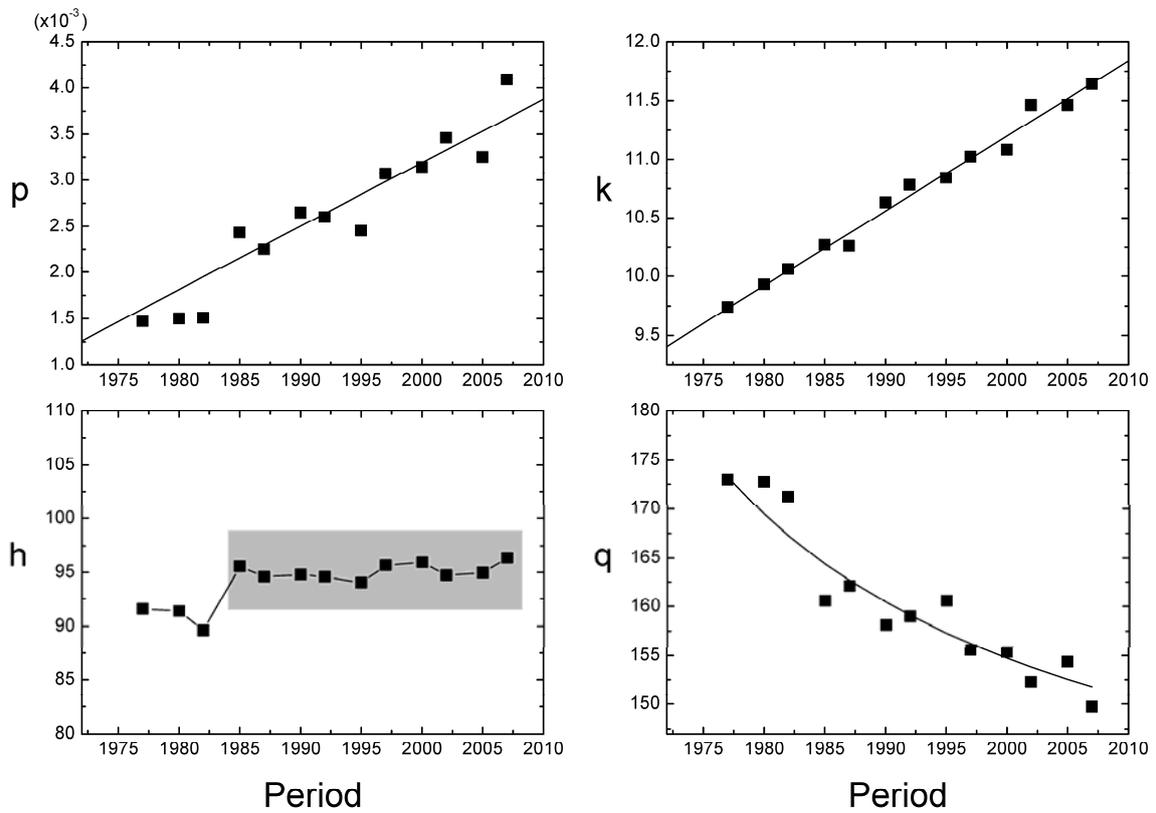

**Figure 2:** The evolution of the quadratic beta function parameters estimated from **Fig. 1**. The p and the k parameters increase linearly by period, while the h parameter does not significantly change, obviously since 1985 (gray area). These evolutions lead to a gradual decrease of the q parameter by period, following the scaling of $(k/p)^{1/2}$.

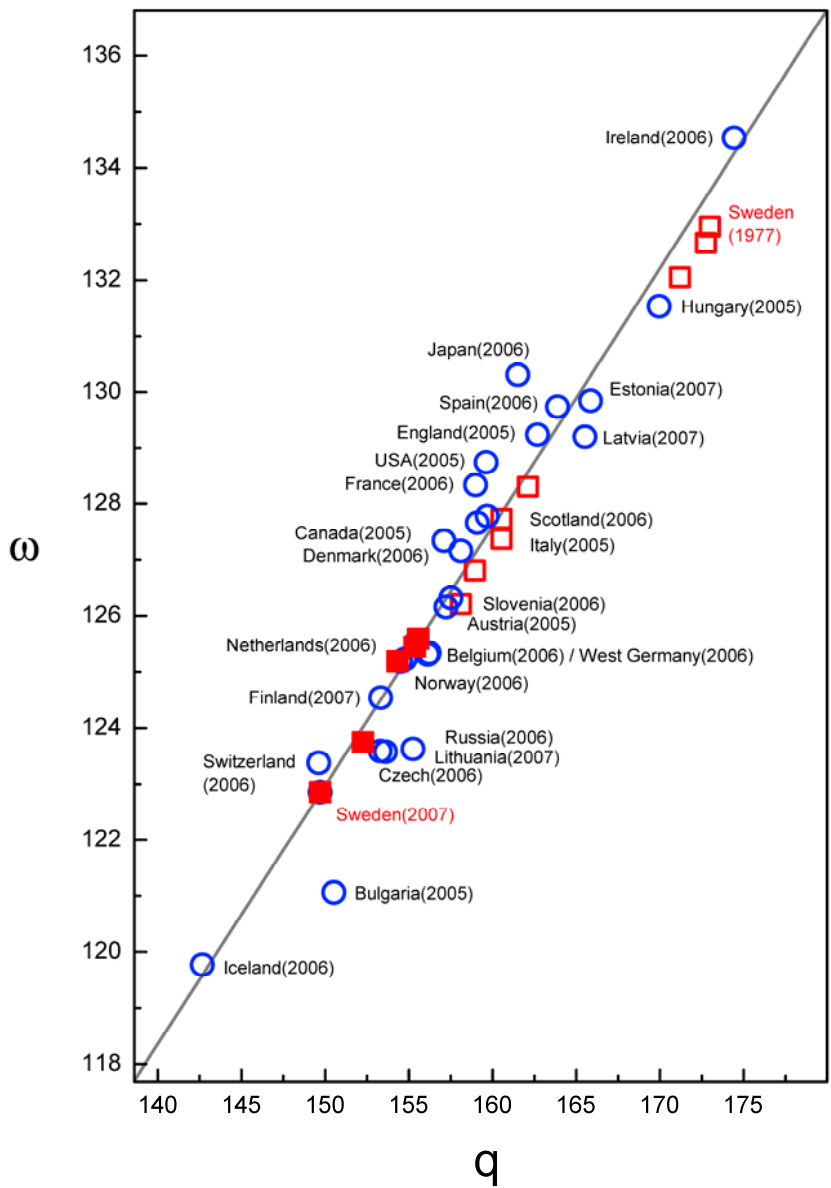

**Figure 3:** The linear relationship of the upper x-intercept q and the theoretical lifespan ω values. The three-decade variation (open and closed squares) for Sweden females is similar to the national variation (circles) for modern industrialized countries. Interestingly, the maximum lifespan ω values linearly decrease with the upper x-intercept q values at a rate of ω = 0.458q + 54.241. The ω values approach ~125 years for Sweden females during the latest decade from 1997 to 2007 (closed squares).

**Table 1. Estimations for Sweden female survival datasets.**

| Datasets | α (yrs) | $\beta_0$ | $\beta_1$ | $\beta_2$ | $r^2$ | h (yrs) | k | q (yrs) | ω (yrs) |
|---|---|---|---|---|---|---|---|---|---|
| 2007 | 88.57223 | 26.32347 | 0.78810 | 0.00409 | 0.99915 | 96.34 | 11.64 | 149.70 | 122.86 |
| 2005 | 88.52680 | 17.85141 | 0.61727 | 0.00325 | 0.99949 | 94.96 | 11.46 | 154.34 | 125.19 |
| 2002 | 87.83450 | 19.58927 | 0.65550 | 0.00346 | 0.99872 | 94.73 | 11.46 | 152.27 | 123.75 |
| 2000 | 87.81336 | 17.81521 | 0.60247 | 0.00314 | 0.99864 | 95.93 | 11.08 | 155.35 | 125.45 |
| 1997 | 87.66873 | 17.06434 | 0.58731 | 0.00307 | 0.99541 | 95.65 | 11.02 | 155.58 | 125.59 |
| 1995 | 87.34920 | 10.82163 | 0.46078 | 0.00245 | 0.99955 | 94.04 | 10.84 | 160.56 | 127.73 |
| 1992 | 86.91140 | 12.47482 | 0.49175 | 0.00260 | 0.99857 | 94.57 | 10.78 | 158.95 | 126.81 |
| 1990 | 86.53163 | 13.17982 | 0.50236 | 0.00265 | 0.99926 | 94.78 | 10.63 | 158.11 | 126.21 |
| 1987 | 86.30050 | 9.860300 | 0.42559 | 0.00225 | 0.99820 | 94.58 | 10.26 | 162.12 | 128.30 |
| 1985 | 85.86844 | 11.91555 | 0.46436 | 0.00243 | 0.99576 | 95.55 | 10.27 | 160.55 | 127.37 |
| 1982 | 85.58855 | 2.065850 | 0.27060 | 0.00151 | 0.99858 | 89.60 | 10.06 | 171.21 | 132.05 |
| 1980 | 85.06656 | 2.594650 | 0.27414 | 0.00150 | 0.99487 | 91.38 | 9.93 | 172.75 | 132.66 |
| 1977 | 84.83856 | 2.584840 | 0.26922 | 0.00147 | 0.99463 | 91.57 | 9.74 | 172.98 | 132.95 |

**Table 2. Estimations for international female survival datasets.**

| Datasets | α (yrs) | r² | h (yrs) | k | q (yrs) | ω (yrs) |
|---|---|---|---|---|---|---|
| Austria (2005) | 88.07717 | 0.99973 | 92.06 | 11.51 | 157.23 | 126.16 |
| Belgium (2006) | 88.17957 | 0.99960 | 95.29 | 11.23 | 154.80 | 125.23 |
| Bulgaria (2005) | 82.71487 | 0.95438 | 94.92 | 9.55 | 150.53 | 121.06 |
| Canada (2005) | 88.93178 | 0.99963 | 100.00 | 10.56 | 157.10 | 127.34 |
| Czech (2006) | 85.58519 | 0.99742 | 94.70 | 10.51 | 153.61 | 123.58 |
| Denmark (2006) | 86.67277 | 0.98848 | 100.58 | 9.96 | 158.11 | 127.16 |
| England (2005) | 87.57264 | 0.99952 | 96.30 | 10.45 | 162.70 | 129.24 |
| Estonia (2007) | 85.47782 | 0.99604 | 95.36 | 9.79 | 165.86 | 129.85 |
| Finland (2007) | 88.76823 | 0.99834 | 93.56 | 11.93 | 153.32 | 124.54 |
| France (2006) | 90.37934 | 0.99963 | 95.20 | 11.72 | 159.00 | 128.33 |
| Germany (2006) | 88.11166 | 0.99869 | 89.77 | 11.92 | 156.22 | 125.34 |
| Hungary (2005) | 83.79004 | 0.99779 | 96.27 | 9.07 | 169.96 | 131.53 |
| Iceland (2006) | 88.78873 | 0.99560 | 99.22 | 11.61 | 142.67 | 119.77 |
| Ireland (2006) | 87.53789 | 0.99960 | 90.75 | 10.44 | 174.44 | 134.53 |
| Italy (2005) | 89.22443 | 0.99977 | 92.76 | 11.62 | 159.09 | 127.66 |
| Japan (2006) | 91.59045 | 0.99942 | 98.17 | 11.47 | 161.51 | 130.31 |
| Latvia (2007) | 83.82117 | 0.99683 | 95.46 | 9.43 | 165.53 | 129.20 |
| Lithuania (2007) | 84.63518 | 0.99738 | 97.72 | 9.82 | 153.30 | 123.59 |
| Netherlands (2006) | 87.85314 | 0.99636 | 95.65 | 11.17 | 155.48 | 125.55 |
| Norway (2006) | 88.49842 | 0.99951 | 94.48 | 11.57 | 154.52 | 125.19 |
| Poland (2007) | 87.63163 | 0.99919 | 91.29 | 11.48 | 156.14 | 125.31 |
| Russia (2006) | 81.29562 | 0.98645 | 98.59 | 8.61 | 155.26 | 123.63 |
| Scotland (2006) | 86.15168 | 0.99884 | 100.50 | 9.49 | 159.68 | 127.78 |
| Slovenia (2006) | 87.42881 | 0.99890 | 94.93 | 10.81 | 157.52 | 126.32 |
| Spain (2006) | 89.47929 | 0.99946 | 89.37 | 11.89 | 163.90 | 129.74 |
| Sweden (2007) | 88.57223 | 0.99915 | 96.34 | 11.64 | 149.70 | 122.86 |
| Switzerland (2006) | 89.69486 | 0.99806 | 97.52 | 11.75 | 149.62 | 123.38 |
| USA (2005) | 87.43102 | 0.99878 | 103.37 | 9.27 | 159.62 | 128.74 |